\documentclass[preprint]{aastex}

\begin{document}

\title{ON FOSSIL DISK MODELS OF ANOMALOUS X-RAY PULSARS}
\author{G.J. Francischelli\footnote{francis@mail.astro.sunysb.edu} \ and
R.A.M.J. Wijers\footnote{rwijers@sbast3.ess.sunysb.edu}}
\affil{Department of Physics and Astronomy, State University of New York}
\affil{Stony Brook, New York 11794}

\begin{abstract}
Currently, two competing models are invoked in order to explain the observable
properties of Anomalous X-ray Pulsars (AXPs).  One model assumes that AXP 
emission is powered by a strongly magnetized neutron star -- i.e., a magnetar.
Other groups have postulated that the unusually long spin periods associated
with AXPs could, instead, be due to accretion.  As there are severe 
observational constraints on any binary accretion model, fossil disk models
have been suggested as a plausible alternative. Here we analyze fossil disk
models of AXPs in some detail, and point out some of their inherent 
inconsistencies.  For example, we find that, unless it has an exceptionally 
high magnetic field strength, a neutron star in a fossil disk cannot be 
observed as an AXP if the disk opacity is dominated by Kramers' law.  However,
standard alpha-disk models show that a Kramers opacity must dominate for the 
case $\log B \ga 12$, making it unlikely that a fossil disk scenario can 
successfully produce AXPs.  Additionally, we find that in order to 
sufficiently spin down a neutron star in a fossil disk, an unusually efficient
propeller torque must be used.  Such torques are inconsistent with
observations of other accreting sytems -- particularly High Mass X-ray 
Binaries.  Thus, our analysis lends credence to the magnetar model of AXPs.

\end{abstract}


\section{INTRODUCTION}
\label{Intro}
Anomalous X-ray Pulsars (AXPs) are a recently discovered subclass of X-ray 
pulsators sharing distinct properties that are markedly different from that
of their high-mass X-ray binary (HMXB) and low-mass X-ray binary (LMXB) 
cousins \citep{mer95, van95}.  In particular, the six known AXPs have an 
extremely narrow range of pulse periods, $\sim 6-12$ s.  This is remarkable
when compared to HMXBs, for example, with spin periods in the range 
$0.069$ -- $\sim 10^4$ s.  Additionally, AXPs have relatively low X-ray 
luminosities, ($L_{\rm x} \sim 10^{34}-10^{36}$ erg s$^{-1}$), no known  
optical companions and are undergoing stable spindown.  They also have 
relatively soft spectra, fit by a combination of power-law and blackbody 
models.  Finally, the neutron stars at the heart of AXPs are thought to be 
young, as evidenced by measured spindown ages in the range $10^3 -10^5$ yr.  
This assumption is enhanced by the observation of several clear supernova 
remnant associations.  A related class of objects (not addressed in this 
paper) are the so-called Soft Gamma Repeaters (SGRs) which share, in their 
quiescent state, similar spin periods, spindown ages, and luminosities.  For a
recent observational review of AXPs, see \citet{mer02}.

The major theoretical hurdle in understanding AXPs lies in discovering the 
source of their X-ray emission.  Rotational power is clearly not responsible
for their power as $\dot{E}_{\rm rot} = I\Omega \dot{\Omega} \ll L_{\rm x}$ 
for typical values.  There have been many attempts to model AXPs in recent 
years but all models can, more or less, be grouped into one of two categories.
The first class of models assume that the source of AXPs are isolated neutron
stars with magnetic field strengths in the range $10^{14} - 10^{15}$ Gauss --
i.e. ``magnetars'' \citep{dun92}.  It was discovered that magnetars could
account for the timing properties of both AXPs and SGRs by simply invoking 
standard magneto-dipole braking \citep{tho95, tho96}.  Furthermore, in this 
model, the source of the X-ray luminosity may be explained by rapid 
magnetic field decay \citep{tho96, col00}.  The second class of models used
to explain the observable features of AXPs invoke some type of accretion. 
Such models do not require neutron stars with unusually strong magnetic fields;
they are assumed to have properties consistent with ordinary radio 
pulsars, $B \la 10^{13}$ G.  In this case, the spindown torque is 
external and, supposedly, a natural consequence of accretion braking.  It was 
quickly shown, however, that traditional accretion models lead to a number of 
problems.   Standard Bondi accretion models \citep{bon52} show that 
unless AXPs all lie near particularly dense regions of the Interstellar Medium
(ISM), for example, near molecular clouds [e.g. \citet{isr94}], it is unlikely
that they are directly accreting from the ISM.  \citet{mer95} suggested that 
AXPs may be members of very low-mass X-ray binaries, (VLMXBs) -- a subclass of
LMXBs with slightly higher magnetic field strengths, $\sim 10^{11}$ G.  One 
possible example of a VLMXB is the X-ray pulsar, $4$U$ 1626-67$. \citet{ver90}
had estimated that a low-mass donor star of about $0.02 M_\sun$ accreting onto
an old neutron star may be responsible for its X-ray emission and, in fact, an
optical identification has been made \citep{cha98} for this system.  Although
this scenario can not be ruled out for AXPs, there are severe observational
restrictions that necessarily limit the possibility of detecting any potential
optical companion they might have \citep{mer98}.  For example, there have been
no detected Doppler modulations in the X-ray frequency for any of the AXPs.  
Such modulations are expected, however, even in the case of accretion from a 
low-mass Helium star or a very low-mass white dwarf.  Additionally, there 
seem to be clear observational differences between the AXPs and 4U $1626-67$.
For example, the LMXB has a much harder spectrum than any of the anomalous
X-ray pulsars.  Another discrepancy is that while 4U $1626-67$ has undergone
clear intervals of both spinup and spindown (to be expected for an accreting
source), this is not the case for the AXPs which show persistent spindown only.
The first non-binary accretion model was suggested by \citet{van95}.  In this 
scenario, AXPs are descendants of close HMXBs and are observed during the post
Thorne-\.{Z}ytkow/spiral-in phase.  In this case the source of the X-ray 
luminosity is a neutron star accreting from a residual disk formed from the 
disrupted companion star.  The apparent youth of AXPs is consistent with this 
model as is their association with supernova remnants.  \citet{gho97} 
suggested that this model can naturally explain the multi-phase spectra of 
AXPs.  For example, disk accretion can account for the power-law phase of the 
spectra whereas a spherically symmetric flow, emanating from the companion's 
envelope would account for the blackbody component.  Of course, this model 
naturally assumes the nascent neutron star can survive the 
common-envelope/spiral-in phase of its evolution.  It has been demonstrated, 
however, that this is not the case \citep{bro95,che93} as hypercritical 
accretion forces $\ge 1 M_\sun$ onto the Thorne-\.{Z}ytkow object, crushing 
it, forming a black hole.  

It is unknown whether or not all of the mass ejected from a core-collapse 
supernova can escape the gravitational well of the embedded compact star.  
With this in mind, it is certainly plausible that a significant amount of mass
may fall back onto a young neutron star after the supernova 
\citep{lin91,che89}.  Most recently, several groups have independently proposed
that fossil disks may interact with young neutron stars and that this
interaction can result in timing signatures similar to what is observed for 
AXPs \citep{cha00,alp01,mar01}.  In this latest challenge to the magnetar 
hypothesis it is  asserted that neutron stars embedded in fossil disks have 
conventional field strengths, similar to those of radio pulsars.  
\citet{cha00} (hereafter CHN) are the first to present a detailed physical 
model of fossil
disk accretion incoporating a time-dependent mass-transfer rate.  In this
model, it is assumed that after a brief transient phase, mass-loss is 
self-similar, obeying a power-law decay.  The narrow range of observed 
spin periods can be explained by assuming the inflow quickly becomes an
advection-dominated accretion flow (ADAF).  \citet{alp01} goes further and
posits that not only can AXPs (and possibly SGRs) be explained by fossil disk
models but varying fossil accretion rates can account for all ``non-standard'' 
young neutron stars that do not appear to be radio pulsars \citep{kas00}.
\citet{mar01} suggest that the environment surrounding AXPs (and SGRs) are 
unusually dense, thereby facilitating the formation of a fossil disk.  This 
so-called ``pushback'' model has been criticized however \citep{dun02} as the
inferred densities calculated by \citet{mar01} very strongly depend on the 
AXP distance, a quantity which is not known with accuracy.  In any case, the 
unifying feature of all these fossil disk models is that the torque causing
the necessary rapid spindown is induced by the propeller-effect \citep{pri72,
ill75,fab75}.  The propeller mechanism allows a rapidly rotating neutron star 
to prevent the flow of material from accreting to its surface.  The resulting 
interaction between the neutron star magnetosphere and the accreting material
somehow causes the neutron star to lose angular momentum.  

Here, we show that all fossil disk models carry inherent inconsistencies.  We
focus our analysis on the CHN model, as it is the only current model to provide
a detailed physical picture of fossil disk accretion but our arguments 
apply generally.  In \S~\ref{fossil}, we show that the final spin period of a 
neutron star embedded in a fossil disk strongly depends on the disk opacity.  
We find that, if free-free and bound-free transitions dominate over scattering
in the disk and $B \la 3.7 \times 10^{13}$ Gauss, the neutron star can not be 
observed as an AXP. Employing a simple alpha-disk model \citep{sha73}, we show
that, in fact, a Kramers opacity must be assumed if one is to believe the 
fossil disk mass-transport rate of CHN.  

In \S~\ref{prop}, we discuss the consequences of the propeller torque model 
employed by CHN.  In order to spin the neutron star down to the observed AXP 
range in a time approximating its spindown age, CHN assumed an ``efficient'' 
propeller torque, allowing the ejected plasma to be spun up to corotation with
the pulsar.  Although the exact mechanism by which a neutron star loses angular
momentum by the propeller effect is not well understood, we show that other
``standard'' torque mechanisms, in fact, yield the opposite result.  As pointed
out by \citet{tho02}, if one, instead, assumes the plasma is ejected with the
specific angular momentum of a particle in a Keplerian orbit at the 
magnetospheric boundary, one must then assume a dipole field several orders of
magnitude greater than what is inferred for radio pulsars in order to yield 
similar results.  In fact, we will show that such propeller mechanisms fail to
produce AXPs.  Of course, the reality of the magnetospheric interaction is a 
very complex magnetohydrodynamical problem so it is hard to say, \emph{a 
priori}, which propeller model is inherently more accurate therefore it is 
necessary to treat all models as competitive and compare the results.  In 
\S~\ref{smcx-1}, we examine the effect of invoking various propeller torques 
to the (hopefully) well-understood HMXB, SMC X-1.  We find the CHN 
``efficient'' propeller suggests a spin period for SMC X-1 that deviates 
orders of magnitude from its observed period.  In \S~\ref{discuss}, we 
summarize our findings and discuss the possible implications.

\section{FOSSIL DISK MODELS}
\label{fossil}
It has long been considered that immediately following the formation of a
neutron star in a core-collapse supernova, some fraction of the ejected plasma
may be unable to escape the gravitational well of the star.  Following
\citet{lin91} and \citet{che89}, CHN suggest a fallback mass $\la 0.1 M_\sun$
is not unreasonable.  At that point, it is unclear what fraction of this 
fallback mass will eventually form the fossil disk but it is suggested that a 
large amount can either be immediately ejected from or accreted onto the 
neutron star.  Comparing this problem with that of the tidal disruption of 
stars by massive black holes, discussed earlier by \citet{can90}, CHN find 
that fallback matter can circularize into an accretion disk in (roughly) a 
local dynamical time.  For their analytical model, CHN assume the dynamical 
timescale is $\sim 1$ ms but claim that the final outcome is insensitive 
to the exact numerical value.  Furthermore, an initial disk mass, $M_0 = 0.006
M_\sun$ is assumed.  Later we will discuss the effects of varying the disk 
mass (as well as the initial spin period of the neutron star) but, for now, we
note that as the observed spin periods of AXPs are in a very narrow range, it 
is not likely possible to vary the initial disk mass much and find results 
consistent with the predictions of CHN.  The newly born neutron star is 
assumed to have ``canonical'' properties with a magnetic field strength in the
range $10^{12} < B < 10^{13}$ Gauss and an initial spin period, $P_0 \sim 15$ 
ms.  Such properties agree with what is observed for the Crab pulsar (PSR 
$0531-21$) although recent analyses have shown that this may not be the norm 
\citep{kas00, alp01}.  

Unlike the case of binary accretion, the fossil disk can not be replenished so
accretion is necessarily a time-dependent phenomenon.  Following 
\citet{can90}, CHN suggest that after the brief transient phase, the 
accretion rate declines self-similarly.  In our analysis, we have devised a 
computer model that can exactly reproduce the analytical model of CHN with the
added benefit that it is simple to vary any parameter and determine its 
influence on the final outcome.  We discuss the results of the model in \S2.1 
and explore the consequences of varying one particular parameter (the opacity)
in \S2.2.

\subsection{Self-Similar Fossil Disks}
\label{self}
Following the arguments of \citet{can90}, CHN suggest that after a dynamical 
time the fossil disk loses mass self-similarly:

\begin{equation}
\label{mdot}
\dot{M} = \left\{
	   	 \begin{array}{ll}
		   \dot{M}_0, & \mbox{$0<t< t_{\rm D}$,} \\
		   \dot{M}_0\left(\frac{t}{t_{\rm D}}\right)^{-\Gamma(\kappa)}, & 
					\mbox{$t \ge t_{\rm D}$}
		 \end{array}
		 \right.
\end{equation}
Here, $t_{\rm D} \sim 10^{-3}$ s, is the local dynamical time and $\Gamma > 1$
is a constant that directly depends on the disk opacity, $\kappa$ (see below).
Not all of the mass lost from the fossil disk will be accreted onto the neutron
star surface.  Thus, in general, the neutron star accretion rate, 
$\dot{M}_{\rm x} \le \dot{M}$ and $\dot{M}_{\rm x}=0$ during the propeller 
phase.  Of course, only the surface accretion rate gives rise to the observed
X-ray luminosity, $L_{\rm x}=GM_{\rm x} \dot{M}_{\rm x}/R_{\rm x}$.
Normalizing to the total initial disk mass, $\dot{M}_0$, it is determined from 
equation~(\ref{mdot}):
\begin{equation}
\label{mdot_0}
\dot{M}_0 = \left(\frac{\Gamma - 1}{\Gamma}\right) \left(\frac{M_0}{t_{\rm D}}
	\right)
\end{equation}
From \citet{can90}, we see the parameter $\Gamma(\kappa)$ directly depends on
the disk opacity.  For a standard parameterization, $\kappa(\rho, T)=\kappa_0 
\rho^{\lambda} T^{\nu}$, where $\rho$ and $T$ are the local density and
temperature respectively, one finds:
\begin{equation}
\label{Gamma}
\Gamma = \frac{38+18\lambda-4\nu}{32+17\lambda-2\nu}
\end{equation}
Thus, a standard Kramers opacity $(\lambda=1,\nu=-7/2)$ yields $\Gamma = 
1.25$ whereas if the opacity is predominantly due to Thomson scattering 
$(\lambda=\nu=0)$, then $\Gamma=19/16 \cong 1.188$.  In their analytic model, 
CHN suggest $\Gamma=7/6 \cong 1.167$.  

The subsequent evolution of the system can be divided into four phases
depending on the relative strengths of several parameters.  The magnetospheric
boundary of a neutron star is often defined as the point where ram pressure of 
infalling matter is balanced by the pulsar's magnetic dipole pressure [see
\citet{fra02} and references therein].  Hence, the flow of accreted matter is
governed by the dipole field for any $r \le R_{\rm m}$.  Here $R_{\rm m}$ is 
the magnetospheric radius which, for a neutron star with $R_{\rm x} = 10$ km 
and $M_{\rm x} = 1.4 M_\sun$, is determined to be:

\begin{equation}
\label{alfven}
R_{\rm m} \cong 0.5 \left(\frac{B^4 R_{\rm x}^{12}}{8GM_{\rm x}\dot{M}^2}
	\right)^{1/7} \approx \ 6.6 \times 10^7 B_{12}^{4/7}\left(
	\frac{\dot{M}}
	{\dot{M}_{\rm Edd}}\right)^{-2/7} \rm {cm}
\end{equation}
In equation (\ref{alfven}), $\dot{M}_{\rm Edd} \approx 9.46 \times 10^{17}$ 
g s$^{-1}$ is the (hydrogen) Eddington accretion rate, $B_{12}=B/10^{12}$ 
Gauss, and the factor of $1/2$ comes from the assumption of a disk geometry.  
Regardless of whether the flow starts out in a disk or has a spherical 
geometry, gravitationally captured plasma can only reach the neutron star's 
magnetosphere if it can fall along a closed field line.  This is only 
permitted within the neutron star's light cylinder, defined such that 
$R_{\rm lc}=c/\Omega=cP/2\pi$.  Thus, plasma can interact with the 
neutron star's magnetosphere if and only if $R_{\rm m}\le R_{\rm lc}$.  If 
this is not the case, the neutron star evolves independently of the fossil 
disk and spins down by magnetodipole braking (i.e. \emph{emitter phase}).  
Once the neutron star spins down sufficiently such that matter can couple to 
its magnetic field, the interaction's effect on the overall spin evolution 
depends on the balance between centrifugal and gravitational accelerations.  
If the pulsar is initially rotating too fast, the neutron star will eject 
infalling plasma, propelling it away tangentially, while simultaneously losing
angular momentum in the process.  This ``propeller'' effect \citep{pri72,
ill75} can be parameterized by a \emph{fastness parameter} $\omega_{\rm s} 
\equiv \Omega/\Omega_{\rm K}(R_{\rm m})$, where $\Omega_{\rm K}(R_{\rm m})=
(GM_{\rm x}/R_{\rm m}^3)^{1/2}$ is the Keplerian angular velocity at the 
magnetospheric boundary.  If the fastness parameter greatly exceeds unity, the 
propeller mechanism is initiated.  For both the propeller and emitter phases,
no material can accrete to the neutron star surface and $L_{\rm x} \propto 
\dot{M}_{\rm x} \sim 0$.  As discussed in the introduction, the way in which
a propelling neutron star loses angular momentum is not well understood.  
However, it is safe to assume that some kind of propeller torque (see 
equation~[\ref{chn_prop}], \S3) acts on the neutron
star, spinning it down with time, and reducing the fastness parameter.  When 
$\omega_{\rm s} \la 1$, surface accretion can take place and, provided the 
transfer rate is sub-Eddington, $\dot{M}=\dot{M}_{\rm x}$.  Contrary to what 
is expected in an X-ray binary, however, $\dot{M}$ steadily decreases in a 
fossil disk and an extended accretion phase never really occurs.  Instead, a 
quasi-equilibrium period, which CHN dub a ``tracking'' phase, occurs, and it 
is during this period that the neutron star can be observed as an X-ray 
pulsar.  Furthermore, they suggest that the tracking phase must be short lived
as the accretion flow quickly becomes advection-dominated (ADAF).  It is 
proposed that an ADAF transition occurs when the accretion luminosity falls to
$\approx 10^{-2}L_{\rm Edd} \approx 1.8 \times 10^{36}$ erg s$^{-1}$ 
\citep{cha00,nar95}.  At this point, it is thought that most of the captured 
matter is ejected prior to reaching the neutron star's surface and, once 
again, $\dot{M}_{\rm x} \sim 0$.  From equation~(\ref{mdot}), the ADAF 
transition time is given by
\begin{equation}
\label{ADAF}
t_{\rm ADAF} \cong t_{\rm d}
	\left(\frac{100 \dot{M}_0}{\dot{M}_{\rm Edd}}\right)^{1/\Gamma} 
\end{equation}
Typically, $t_{\rm ADAF} \sim 20,000-40,000$ yr.  In sum, the neutron star can
only be observed as a bright X-ray source during the tracking phase of its 
evolution, $t_{\rm track} \la t \la 2t_{\rm ADAF}$.

To determine the spin evolution of the neutron star + fossil disk system, a 
spinup/spindown torque is needed.  For $R_{\rm x} < R_{\rm m} < R_{\rm lc}$, 
CHN propose the following \citep{men99}:
\begin{equation}
\label{chn_prop}
\dot{J}=I\dot{\Omega}=2\dot{M}R_{\rm m}^2\Omega_{\rm K}(R_{\rm m})\left[
	1-\frac{\Omega}{\Omega_{\rm K}(R_{\rm m})}\right]
\end{equation}
Assuming an arbitrary value for $\Gamma$, equations~(\ref{mdot}) 
and~(\ref{chn_prop}) can be combined to yield an analytic formula for 
$\Omega(t)$, in terms of incomplete gamma functions.  For the particular case
$\Gamma=7/6$, CHN find a solution using exponential integrals (see their eq. 
4).  Instead of using their analytic method, however, we solved the above 
equations numerically, using an Eulerian integration scheme.  The results 
of our analysis (with $\Gamma=7/6)$ is shown in Figure 1.  Here we plot the 
spin evolution of the neutron star as a function of magnetic field strength 
until the approximate onset of the ADAF phase, $2t_{\rm ADAF} \sim 38,000$ yr.
Initial conditions $P_0 = 15$ ms and $M_0 = 0.006 M_\sun$ are assumed.  We 
see that a high magnetic field strength is needed in order for the neutron 
star to be observed as an AXP.  For the above initial conditions, neutron 
stars with $B_{12} \la 3.9$ do not spin down efficiently and, consequently, 
are observed as radio pulsars.  On the other hand, for $B_{12} \ga 3.9$, the 
neutron star, after a brief emitter phase, undergoes a rapid propeller 
spindown and enters the tracking phase of its evolution at $t_{\rm track} \sim
10^4$ yr.  Our numerical analysis agrees with the analytic model of CHN (see 
their figure 1).  

The initial spin period of neutron stars is largely unknown and the 
long-accepted paradigm that assumed the Crab pulsar is the prototypical young 
neutron star is now being challenged \citep{kas00}.  Thus, it is important to 
investigate how the previous results depend on $P_0$.  In fact, we find that 
the duration of the AXP phase does not strongly depend on the initial period.
For example, assuming $P_0 = 150$ ms had little effect on the final 
evolutionary state of the system (pre-ADAF).  For a particular case, consider 
a neutron star with $B_{12} = 7.5$ and $P_0 = 150$ ms.  Our analysis has shown
that it is able to spin down via the propeller effect to the AXP range $(10.0$
s$)$ in a time $t_{\rm ADAF} \cong  4 \times 10^4$ yr.  The only deviation 
from what is seen in figure 1 is in the early evolution.  In this case, we 
found a short-lived accretion phase can occur $\sim 100$ yr after birth.  
This is compensated by a somewhat steeper propeller cycle and a slightly 
shorter tracking phase.  The overall evolution, however, is qualitatively the 
same as the $15$ ms case and, as this case is representative of the general 
trend, we therefore conclude that the CHN model can accomodate a wide range of
initial periods.  

Another free parameter to be considered is the initial disk mass.  Even 
assuming a fallback mass $\la 0.1 M_\sun$, it is unclear how much matter will 
circularize into an accretion disk.  In the initial period following the 
supernova, much of this material may be accreted onto the neutron star, 
possibly hypercritically.  Additionally, as detailed modelling probably 
depends on the progenitor history, we make only a few qualitative remarks 
about assigning possible values to the disk mass.  As CHN have shown, an AXP 
can be formed even if $M_0 \ll 0.1 M_\sun$.  However, there still must be a 
sufficient amount of disk matter in order to penetrate the neutron star's 
light cylinder.  From equation~(\ref{alfven}), we see the magnetospheric radius
increases as the accretion rate falls and if $\dot{M}$ falls below 
some critical value, matter will be unable to couple to the closed field lines
and no (extended) propeller phase will take place.  Our numerical simulations 
indicate that $M_0 \ga 7.5 \times 10^{-4} M_\sun$ is needed for the 
neutron star to be observed as an AXP.  A neutron star embedded in a fossil
disk with significantly less mass than this, rarely leaves the emitter phase
of its evolution and, hence, would be observed as a radio pulsar.  The exact 
critical value strongly depends on the neutron star's magnetic field strength;
for $B_{12} \ga 7.5$, even more disk matter is needed.  The most important 
observational constraint in determining the initial disk mass is the narrow 
distribution of AXP spin periods.  Deviations far from the CHN sample value of
$0.006 M_\sun$, although often producing neutron stars with a tracking period,
often yield a wider range of spin periods than what is currently observed.  
Thus, for our model, $M_0 = 0.006 M_\sun$ was consistently used.

\subsection{The Effect of Varying Disk Opacity}
\label{opacity}

One quantity that deserves special attention is the disk opacity, parameterized
by $\Gamma(\kappa)$ in equation~(\ref{Gamma}).  We now show that if free-free
and bound-free transitions dominate over electron scattering within the disk 
(i.e., if a Kramers opacity is assumed), then a canonical neutron star will 
never reach a tracking phase and, consequently, will not be observed as an 
AXP.  We recall from equation~(\ref{Gamma}) that $\Gamma_{\rm es}=19/16=
1.1875$ and $\Gamma_{\rm Kr} = 1.25$ so it is not immediately obvious that 
such numerically similar parameters could lead to completely different 
evolutionary scenarios.  Figure 2, however, illustrates a strong correlation
between $\Gamma$ and the overall spin evolution of the neutron star.  Here we
plot the final spin period of a neutron star embedded in a fossil disk as a 
function of $\Gamma(\kappa)$ for various magnetic field strengths.  An 
efficient propeller torque given by equation~(\ref{chn_prop}) is assumed.  As 
we have seen in section \S~\ref{self}, the system's evolution also depends on 
the neutron star's magnetic field.  In particular, for a given disk opacity, 
the field strength determines the extent of the propeller phase and, 
ultimately, whether or not a tracking phase can occur at all.  This dependence
is shown in figure 2 by comparing opacity-period evolution curves for 
$B_{12} =2.5$, $5.0$, $7.5$, $10.0$ and $40.0$.  Although the shape and 
qualitative behavior are the same for the various field strengths, the overall
range of final periods differs greatly.  Generally, each curve shows 
essentially no dependence on opacity below some cutoff value.  As $\Gamma$
increases past a value, $\Gamma_{\rm c}$, however, there is a sharp drop in 
final spin period followed by a more gradual decline. These curves illustrate
the fact that, depending on the strength of the field, a neutron star embedded
in a disk with opacity parameter greater than some $\Gamma_{\rm c}$, will not
be able to spin down to the AXP range and, consequently, it will end its 
observable life as a radio pulsar.  The decline in final spin period with 
increasing $\Gamma$, can be interpreted physically by noting from 
equation~(\ref{mdot}) that the overall mass-transport rate steeply decreases 
with time by a factor $\propto t^{-\Gamma}$ and for $\Gamma > \Gamma_{\rm c}$,
disk matter is depleted too quickly to sustain an extended propeller phase.  A
secondary cause for the decline is the ADAF transition, 
and subsequent X-ray shut-off, must necessarily occur at earlier times for 
increased $\Gamma$.  For $\Gamma = 7/6$, our results agree with the conclusions
of CHN.  As mentioned previously, in this case, the minimum field strength
needed to support a tracking phase $\sim 3.9 \times 10^{12}$ G.  Thus, in 
figure 2, $\Gamma_{\rm CHN} < \Gamma_{\rm c}$ for all fields except $B_{12}=
2.5$.  If the fossil disk were to be dominated by electron scattering, we find
that a stronger field is needed in order for the neutron star to enter a 
tracking period before the ADAF phase of its evolution.  In this case, the 
minimum field required is $\sim 6.6 \times 10^{12}$ Gauss.  For $B_{12}=5.0$, 
$\Gamma_{\rm es} > \Gamma_{\rm c}$ and for $B_{12}=7.5$, $\Gamma_{\rm es} \sim
\Gamma_{\rm c}$.  If the disk is dominated by a Kramers opacity, however, 
unusually strong magnetic fields are necessary in order for the neutron star
to be observed as an AXP.  In fact, $\Gamma_{\rm Kr} \gg \Gamma_{\rm c}$ for 
all field strengths below $\sim 3 \times 10^{13}$ Gauss.  All neutron stars
with magnetic fields below $B_{\rm Kr} \sim 3.7 \times 10^{13}$ Gauss and 
otherwise similar initial conditions will remain a radio pulsar throughout its
observable lifetime.  Of course, the strength of this conclusion directly 
depends on the exact point at which the ADAF transition commences.  As this 
quantity is not known with accuracy, the true value of $B_{\rm Kr}$ may be
either higher or lower.

We have seen that the fossil disk opacity is a key parameter in determining 
the ultimate evolution of the neutron star.  In particular, we have shown that
if a Kramers opacity dominates, it is unlikely (though not impossible) for an 
AXP to be observed whereas if electron scattering is more important, it is 
more likely that a ``canonical'' neutron star can become an AXP.  In fact,
we now show that, for $B_{12} \ga 1$, a Kramers opacity must be assumed.
In order to calculate the opacity we adopt the so-called alpha-prescription of
\citet{sha73}, implicitly assuming that a thin-disk structure is valid for 
fossil disks.  Such an assumption seems valid as  it is probably reasonable to
expect the fossil disk is approximately thin in the regime, $t_{\rm dyn} \ll t 
< t_{\rm ADAF}$.  Furthermore, we assume the (kinematic) turbulent 
viscosity, $\nu$, can be parameterized according to the standard relation, 
$\nu \equiv \alpha c_{\rm s} h$.  Here, $c_{\rm s}$ is the local (isothermal) 
sound speed and $h$ represents the disk's vertical scale-height.  It has been 
argued on physical grounds that the viscosity parameter, $\alpha < 1$ 
\citep{sha73}.  The thin-disk assumption is parameterized by the condition 
$h/r \sim c_{\rm s}/v_{\phi} \ll 1$, where $r$ is the radial measure of the 
disk and $v_{\phi}(r) = r \Omega_{\rm K}(r)=\sqrt{GM_{\rm x}/r}$ represents 
the azimuthal flow within the disk.  The general procedure for determining the 
time-evolution of a viscous disk undergoing mass-transfer variations has been
extensively studied.  \citet{bat81}, for example, have shown that the
time-dependent conservation equations lead to a non-linear diffusion equation
for the surface density, $\Sigma(r,t)$.  It has been shown, however, 
\citep{fra92}, that if at each point in time, and for a given $\dot{M}(t)$, the
timescale in which matter diffuses through the disk greatly exceeds the 
dynamical timescale, $\sim r/v_{\phi}(r)$, the disk may be considered to be 
approximately steady-state.  Except at early times, this condition is generally
valid for the fossil disk we are considering and, thus, in our analysis, we 
assume steady-state solutions may be used for each value of $t$.  This approach
is roughly equivalent to numerically integrating the diffusion equation of 
\citet{bat81}.

We first assume an opacity dominated by bound-free transitions, 
$\kappa_{\rm Kr} \cong 4 \times 10^{25} Z (1+X) \rho T^{-7/2}$ cm$^2$ 
g$^{-1}$, \citep{sch58} where $X \approx 0$ and $Z \la 1$ are the hydrogen and 
heavy element mass fractions, respectively.  Following the analysis of 
\citet{fra92}, we can then estimate the region within the disk where a 
Kramers-type opacity may be dominant.
\begin{equation}
\label{kramers}
\kappa_{\rm Kr}(r,t) = 90 [Z(1+X)]^{1/2} \dot{M}_{17}^{-1/2}(t) r_{10}^{3/4} 
f^{-2}(r) \rm {\ cm}^2{\rm g}^{-1}
\end{equation}
Here $\dot{M}_{17}=\dot{M}/10^{17}$ g s$^{-1}$, and $r_{10}=r/10^{10}$ cm as 
usual and $f(r) \equiv [1-(r_{\rm i}/r)^{1/2}]^{1/4}$.  Fortunately, the 
result is independent of $\alpha$ and, thus, does not depend on the poorly
understood viscosity mechanism within the disk.  The quantity, 
$r_{\rm i}$ represents the inner edge of the disk and for neutron stars, 
$r_{\rm i} = R_{\rm m}$, given by equation~(\ref{alfven}).  In our analysis, 
we ignore the perturbations brought about by edge effects and concentrate on 
the region, $r \gg r_{\rm i}$, $f(r)\approx 1$.  At temperatures $\ga 10^4$ K,
the other dominant source of opacity is electron scattering, $\kappa_{\rm es}=
0.2 (1+X)$ cm$^2$ g$^{-1}$.  From equation~(\ref{kramers}), we can estimate 
the region of the fossil disk where 
$\kappa_{\rm Kr} \ga \kappa_{\rm es}$.
\begin{equation}
\label{kram_rad}
r_{\rm Kr}(t) \ga \left(3 \times 10^6\right) \left[\frac{1+X}{Z}\right]^{2/3}
\dot{M}_{17}^{2/3}(t) \rm{\ cm}
\end{equation}
A $1.4 M_\sun$ neutron star is assumed throughout our analysis.  From
equation~(\ref{alfven}), we estimate the magnetospheric boundary of the neutron
star, $R_{\rm m}$, as the point at which the disk truncates.  We use this 
result along with equation~(\ref{kram_rad}) to determine the conditions 
necessary such that electron scattering may be ignored.  I.e., we calculate 
the necessary conditions such that $R_{\rm m}(t) \ge r_{\rm Kr}(t)$.  
\begin{equation}
\label{mdot_no_es}
\dot{M}_{17}(t) \la 50 \left[\frac{1+X}{Z}\right]^{-7/10} B_{12}^{3/5}
\end{equation}
Finally, combining the above relation with the mass-transfer 
equation,~(\ref{mdot}), we can estimate the time after which electron 
scattering may be completely ignored within the disk.  
From~(\ref{mdot}),~(\ref{mdot_0}), and~(\ref{mdot_no_es}), we find
\begin{equation}
\label{kram_time}
t_{\rm Kr} \ga 20 \left[\frac{1+X}{Z}\right]^{14/25} B_{12}^{-12/25} \rm{\ yr}
\end{equation}
For the CHN AXP-cutoff $B_{12} \sim 4$, this corresponds to a value of $\sim 
10$ yr.  With $B_{12} \sim 37$, $t_{\rm Kr} \sim 3.5$ yr.  In any case, we
see that, regardless of the magnetic field strength, the fossil disk model
of CHN must be Kramers-dominated throughout its evolution and $\Gamma = 1.25$ 
is the proper choice of exponents in equation~(\ref{mdot}).  Coupling this 
with the preceding analysis and figure 2, we have now shown that, unless the 
neutron star has an extraordinarily high magnetic field, in excess of $\sim 
3.7 \times 10^{13}$ Gauss, the CHN model fails to produce an anomalous X-ray 
pulsar. 

\section{PROPELLER TORQUES}
\label{prop}
In its most general form, the exact manner in which angular momentum is 
transferred between an accretion disk and a neutron star is a complex 
magnetohydrodynamical problem to which no simple analytic solution  
has been determined.  More than two decades of numerical analyses have 
resulted in only minor modifications to the original model of \citet{gho79}
[but see \citet{wan87}].  It has been determined that when modelling disk 
accretion, it is important to include the effects of magnetic torques within 
the disk to the overall spin rate.  Ghosh \& Lamb were able to show that for 
slow rotators $[\Omega/\Omega_{\rm K}(R_{\rm m}) \equiv \omega_{\rm s} \ll 
1]$, magnetic coupling may enhance spin-up torques by as much as 40\%.  For 
$\omega_{\rm s} \la 1$, the opposite is true, and magnetic effects might 
actually oppose the spin-up.  However, as we have seen, due to a 
time-dependent mass-transfer rate, a neutron star in a fossil disk never 
really has an extensive accretion period nor can it reach true equilibrium.  
Unfortunately, unlike the case for accretion, angular-momentum transfer during
the propeller phase is not well modelled at all.  At present, no \emph{ab
initio} theory exists to compute the torque from an accretion disk on a 
magnetized star.  However, over time, several approximate schemes have been
introduced, generally based on basic physical principles such as conservation
laws \citep{pri72, ill75, fab75}.  Additionally, \citet{men99} have introduced
the ``efficient'' propeller torque, given by equation~(\ref{chn_prop}).  As
several competing models exist, the best we can do is compare the models with 
each other and, most importantly, with observations.  With this in mind, it is
important to emphasize that while we can not really claim one model to be 
intrinsically ``better'' than another, we \emph{can} reject a particular model
if observations force our hand.

For a strong propeller, $\omega_{\rm s} \gg 1$, and equation~(\ref{chn_prop})
can be written in the form:
\begin{equation}
\label{chn_torque}
N_{\rm CHN}=I\dot{\Omega} \cong -2 \dot{M} R_{\rm m}^2 \Omega
\end{equation}
Thus, we see the CHN model assumes a rapidly rotating neutron star 
propels matter from the magnetospheric boundary with (twice) the angular 
velocity of the neutron star, thereby spinning it up to corotation.  
Historically, other propeller models that have been used are not quite so 
efficient.  In fact, it has generally been assumed that the propelled matter 
is, instead, ejected from the magnetospheric boundary with the specific 
angular momentum of a particle in an escaping parabolic orbit.
\begin{equation}
\label{j_prop}
N_{\rm J}=I \dot{\Omega}=-\dot{M}R_{\rm m} v_{\rm esc}(R_{\rm m}) = -\sqrt{2}
\dot{M}R_{\rm m}^2 \Omega_{\rm K}(R_{\rm m})
\end{equation}
The J-subscript in equation~(\ref{j_prop}) indicates we have used angular 
momentum methods to estimate the propeller torque.  An alternative approach
is to employ energy conservation principles.  \citet{fab75} has suggested 
that, over time, the rotational kinetic energy of the neutron star will be 
transmitted through shocks to the wind plasma falling near the magnetospheric 
boundary. Consequently, the infalling gas will heat up and be dispersed when
it attains escape velocity.  Thus, we find $\dot{E}=I\Omega \dot{\Omega}=-
\frac{1}{2}\dot{M}v^2_{\rm esc}(R_{\rm m})=-G \dot{M}M_{\rm x}/R_{\rm m}$.  
The propeller torque is therefore expressed as
\begin{equation}
\label{e_prop}
N_{\rm E}=I \dot{\Omega}=-\frac{\dot{M}R_{\rm m}^2 \Omega_{\rm K}(R_{\rm m})}
{\omega_{\rm s}}
\end{equation}
where $\omega_{\rm s}$ is the fastness parameter.  Upon comparison of
equations~(\ref{chn_torque}) -~(\ref{e_prop}), we find the relationship between
the various proposed propeller torques:
\begin{equation}
N_{\rm CHN}=\sqrt{2}\omega_{\rm s}N_{\rm J}=2\omega_{\rm s}^2 N_{\rm E}
\end{equation}
In the evolution models just discussed here, we found that the fastness 
parameter varies in the
range $\sim 20-75$.  It is therefore not very surprising that the choice of 
propeller torques has a great deal of influence on the overall evolution of 
the neutron star/fossil disk system.  This is illustrated in figure 3.  Here,
we plot a period-time relationship for the various propeller models, keeping 
all other initial parameters fixed.  We have assumed, as usual, an initial
fossil disk mass of $M_0 = 0.006M_\sun$ and the initial field strength and spin
period are $7.5 \times 10^{12}$ Gauss and $15$ ms, respectively.  In order to 
make
our argument independent of the results of the preceding section, we have kept
the opacity parameter fixed at $\Gamma(\kappa) = 7/6$, the original CHN
value.  Clearly, employing the efficient propeller torque of \citet{cha00} 
makes all the difference in determining the overall evolution.  When the torque
given by equation~(\ref{chn_torque}) is used, the neutron star, after a sharp
propeller cycle reaches a tracking phase at $\sim 1.5 \times 10^4$ yr.  It is
then visible as a bright AXP until $2t_{\rm ADAF} \sim 3.8 \times 10^4$ yr, at
which point its final spin period is $13.5$ sec.  For both the energy and
angular momentum propellers, the neutron star never reaches equilibrium and 
despite much higher fastness parameters $(\sim 70)$, the neutron star's final 
period never exceeds $\sim 0.2$ s.  Note the early evolution for the energy
and angular momentum propellers deviate slightly as the energy propller, for 
these initial conditions, lacks an early emitter phase, and begins propelling
matter near $t \sim 0$.  At later times, the evolution traces the angular 
momentum propeller case exactly.  Of course, if given sufficient time, we 
would see the angular momentum propeller force much higher spin periods but 
only at times $t \gg 2t_{\rm ADAF}$.  Thus, neither the angular momentum nor 
the energy propellers yield AXPs but, instead, ordinary radio pulsars.  

We recall, from \S~\ref{opacity}, that the choice of $\Gamma(\kappa)$ strongly
determines the evolution of a neutron star in a fossil disk provided the CHN 
torque is employed.  In fact, we found that exceptional magnetic fields are 
required if one were to assume a physically realistic Kramers opacity.  The 
exact value of $\Gamma$ is much less important to the overall evolution for 
the other propeller torques, however.  Assuming a disk dominated
by electron scattering, for example, does not change the final spin period for
either the angular momentum or energy propellers.  For a Kramers disk, 
$(\Gamma = 1.25)$, we also see no early evolutionary changes in spin period.
However, in this case, the onset of ADAF flow occurs much earlier 
$(t_{\rm ADAF} \cong 2650$ yr by equation~[\ref{ADAF}]).  For a Kramers disk,
therefore, we find the final spin period for \emph{all three} propeller torques
is the same: about $0.13$ s. As pointed out by \citet{tho02}, increasing the
magnetic field strength by an order of magnitude does allow the less efficient
propller torques to spin down the neutron star to longer periods, but we find
that, regardless of the strength of $B$, the J-propeller and E-propeller can 
never produce an anomalous X-ray pulsar.  For example, in order to spin down
a neutron star in a fossil disk by the angular momentum propeller torque to a 
final period of $\sim 5$ s, it is necessary to increase the initial field 
strength to $\cong 9.8 \times 10^{13}$ G.  In fact, even at such high fields, 
the neutron can never reach equilibrium.  Instead, as $\omega_{\rm s} \ga
24$ (and is slowly decreasing) at $t \sim 2t_{\rm ADAF}$, there will be no
tracking period at all.  Borrowing from the terminology of \citet{cha00}, we 
dub such a neutron star a ``dim propeller.''  Spun down into the graveyard
$(P_{\rm f} \sim 5.3$ s$)$, the neutron star is a weak emitter in both radio
and X-rays.  This trend is even more obvious for the case of the energy 
propeller.  Here, in order to spin the neutron star down to a final 
period of $5.9$ s, a magnetic field $\ga 8.6 \times 10^{14}$ G is needed.
Even ignoring the fact that such a field strength is already well into the 
so-called ``magnetar'' range, we find this neutron star, too, is destined to be
a dim propeller $[\omega_{\rm s}(2t_{\rm ADAF}) \ga 140]$.  Thus, we see the 
results of \citet{cha00}, namely that AXPs can be produced from ordinary radio
pulsars, hinges upon the condition of an efficient propeller torque.  If one 
can show that a canonical neutron star can \emph{not} spin propelled matter at
the magnetospheric boundary to corotation, then it is a necessary consequence
that fossil disk models, as outlined by \citet{cha00}, cannot produce 
anomalous X-ray pulsars.

\section{A COMPARISON OF VARIOUS PROPELLER TORQUES FOR SMC X-1}
\label{smcx-1}
Initially discovered during a rocket flight \citep{pri71} and identified by
\emph{Uhuru} as a discrete X-ray source in the Small Magellanic Cloud 
\citep{leo71}, SMC X-1 is the most luminous and one of the most extensively 
studied of the X-ray pulsars.  Its binary nature, implied by periodic X-ray
eclipses \citep{sch72} was confirmed by the identification of an optical
companion: B0I supergiant, Sk 160 \citep{web72,lil73}.  The 3.89 day orbital
period is decaying at a rate $\dot{P}_{\rm orb}/P_{\rm orb} = (-3.36 \pm 0.02)
\times 10^{-6}$ yr$^{-1}$ \citep{lev93}, presumably due to tidal interactions
between the orbital motion and the primary's rotation.  The orbital elements of
the system have been measured with the neutron star's mass estimated at
$M_2 \cong 0.8 - 1.8 M_\sun$ and the primary mass and radius 
given by $M_1 \cong 19 M_\sun$ and $R_1 \cong 18 R_\sun$ respective 
\citep{pri77}.  For consistency, we will continue to assume $M_2 \equiv 1.4
M_\sun$ and $R_2 \equiv 10^6$ cm.  The orbit is circular [$e < 0.00004$, 
\citep{lev93}] and Kepler's third law suggests an orbital separation $a \sim
30 R_\sun$ .  With an 0.71 s pulse period \citep{luc76}, SMC X-1 is the
only X-ray pulsar undergoing stable spinup at a rate $\dot{P} \cong -1.2 \times
10^{-11}$ s s$^{-1}$ \citep{kun93, kah99}.  Spinup implies the neutron star is 
accreting and, indeed, there is evidence of an accretion disk around SMC X-1
\citep{van77,tje86}.  Its luminosity, $L_{\rm x}$, has been measured to vary 
from $\sim 10^{37}$ erg s$^{-1}$ in the low-intensity state to the 
extremely high value of $5 \times 10^{38}$ erg s$^{-1} \sim 5 L_{\rm Edd}$ in 
the high-intensity state.  The presence of an accretion disk, excess X-ray 
luminosity and persistent spinup, together suggest mass-transfer may be due to
atmospheric Roche-lobe overflow from the massive primary \citep{sav79}.  
Recently, an X-ray burst was discovered from SMC X-1 \citep{ang91}, which was
suggested to be due to instabilities in the accretion flow, i.e., a type II
burst \citep{li97}.  If, indeed, SMC X-1 is a member of the class of 
``bursting pulsars,'' like the recently discovered transient X-ray pulsar,
GRO J1744-28, then a diminished pulsar magnetic field might be the cause of 
such bursts.  In fact, \citet{li97} have estimated a magnetic moment for SMC 
X-1 to be somewhat lower than normal, $\mu = B R^3_2 \sim 10^{29}$ G cm$^3$.
This is consistent with claims that magnetic field decay is connected with
binary accretion [see \citet{fra02} for a discussion].

In order to test the various propeller torques of the preceding section, we 
apply a modified accretion model to the SMC X-1/Sk 160 system.  Since the 
giant phase is only about $\sim 10$\% of the main sequence lifetime, it is
important to consider the mass transfer mechanism for the progenitor 
system as well.  Most likely, the progenitor to the supergiant primary is an 
09 V star, with ZAMS mass and radius, $\sim 20 M_\sun$ and $\sim 10
R_\sun$.  Its main sequence lifetime can be estimated by the mass-luminosity 
relation, $L \propto M^{3.5}$, such that $\tau_{\rm ms} \sim 10^{10}$ yr 
$(M/M_\sun)^{-2.5} \cong 5.6 \times 10^6$ yr.  We assume the progenitor lies 
within its Roche lobe throughout the hydgrogen core-burning phase so 
mass transfer is, then,  most likely due to a steady stellar wind, of order 
$\sim 10^{-7} M_\sun$ yr$^{-1}$ \citep{kud00}.  Thus, the overall evolution of
the system has two distinct parts:  In the first part, an O9 V  star transfers
mass, by steady stellar wind, to a young neutron star with initial spin and 
field strength, $\sim 15$ ms and $\sim  10^{13}$ G.  During this phase
of the evolution, mass transfer is highly non-conservative as only a fraction
of the mass emitted by the primary can be captured by the neutron star, 
$\dot{M}_2 
= -f_{\rm c} \dot{M}_1$.  The non-conservative phase lasts the entire 
main-sequence lifetime of the primary.  In the second part of the evolution, a
B0I supergiant conservatively transfers mass via Roche-lobe overflow 
(while non-conservative wind-transfer continues) for a 
time $ \sim 0.1 \tau_{\rm MS} \sim 6 \times 10^5$ yr.  Accretion-induced field
decay is assumed for both parts of the evolution.  The evolution code for 
such a model, while not much different from the original code for the fossil 
disk model described above, is almost identical to the analysis described in 
\citet{fra02}.  Thus, for details of the input physics assumed, we refer the 
interested reader there.  For the sake of brevity, here we simply show the 
results of our analysis in figure 4.  We plot the neutron star spin period
as a function of time for all three propeller models.   Of course, 
we have no way of estimating how long SMC X-1 has been undergoing shell 
burning so the most we can do is compare the known spin period of $0.71$ s 
to the range of predicted values calculated during the supergiant phase.

If an energy propeller torque is assumed we see that, after a brief emitter 
phase, the neutron star in SMC X-1 undergoes an extended propeller cycle.  
Weak propeller-torquing allows the neutron star to gradually spin down until 
it reaches a maximum period of about $30$ s at $t \sim 3.1$ Myr.  At this 
point, accretion spinup commences and continues until a renewed 
propeller-spindown at $t \sim 4.4$ Myr. This leads to a more gradual 
accretion cycle and, at the onset of hydrogen-shell burning, at $5.6$ Myr, the
neutron star is still in the graveyard at $P = 5.3$ s.  Overall, during the 
main-sequence evolution, the radiation-driven wind can only transfer $3.6 
\times 10^{-4} M_\sun$ to the neutron star surface.  As soon as the primary 
overflows its Roche-lobe, however, heavy accretion ensues, recycling the 
neutron star back into observability, spinning it up and bringing down its 
magnetic field.  After a total evolution time of $6.2$ Myr, $P_{\rm f} = 0.99$
s.  Employing the accretion-induced field-decay model of \citet{fra02}, the 
final magnetic field is calculated to be $B_{\rm f} \sim 4 \times 10^{10}$ G.  
For the case of an angular momentum (J) propeller, propeller-spindown is much
more rapid and occurs much earlier in the system's evolution.  As a result,
the accretion phase also occurs much earlier and the pulsar spins back up to 
$\sim 2.2$ s at the end of the core-hydrogen burning phase.  Once again, a 
rapid accretion cycle occurs during Roche-lobe overflow and the final spin 
period is $0.52$ s. As the total accretion time is longer, we find the 
final magnetic field is necessarily lower for the angular momentum propeller
and $B_{\rm f} = 1.3 \times 10^{10}$ G.  Finally, we look at the case of the
efficient, CHN propeller torque.  Here, propeller spindown is even more rapid 
than before and the neutron star reaches $P = 31.3$ s in only $\sim
3.3 \times 10^4$ yr. Unlike what was observed for the other two propeller 
models, here there is no subsequent recycling and the neutron star remains,
roughly, in spin equilibrium for the duration of the main-sequence evolution.
Consequently, no mass is accreted to the neutron star surface during this
period.  As soon as the primary begins to overflow its Roche lobe, however, a 
new propeller cycle spins the neutron star down further into the graveyard,
reaching $P_{\rm max} \sim 100$ s.  Finally accretion is allowed, but the 
pulsar can never be sufficiently recycled back into observability and 
$P_{\rm f} = 15.9$ s.  Altogether, only $9.7 \times 10^{-5} M_\sun$ is accreted
to the neutron star surface and, as a consequence of such weak accretion,
$B_{\rm f} \cong 1.2 \times 10^{12}$ G.  

In conclusion, we see that both the energy and angular momentum propellers,
for the given initial conditions, yield reasonable spin histories for our model
of SMC X-1.  The magnetic field evolution is not as accurate, however, as 
the heuristic field-decay model of \citet{fra02}, while yielding results 
consistent with observations of LMXBs and relativistic binary pulsars, seems
to deviate by a factor or $\sim 2.5 - 10$ from the magnetic moment 
calculations of \citet{li97}.  Although for the purpose of this paper, we are
only interested in the spin evolution of neutron stars for the various
propeller torques, the larger subject of pulsar magnetic field decay in 
general remains an open issue.  The most important result of our analysis, is 
that we have definitively shown that the CHN propeller torque model is 
completely inconsistent with observations of SMC X-1.  In particular, we have 
shown that such a propeller torque yields a neutron star with a spin period 
somewhere in the range $15.9 - 100$ s, orders of magnitude different from
the observed period.  On this basis, we reject the efficient propeller torque 
model and determine it is extremely unlikely that a neutron star can propel 
matter to corotation.  Since we have also shown that less efficient (yet more 
physically plausible) propeller torques can not spin neutron stars in fossil
disks down into the AXP range, regardless of their magnetic field strength, we 
must reject fossil disk models of anomalous X-ray pulsars.

\section{DISCUSSION}
\label{discuss}

Accretion from a fossil disk is the latest scenario that has been suggested as 
a plausible alternative to ultramagnetized neutron stars, or magnetars, as a 
way of explaining the timing signatures of anomalous X-ray pulsars.  Although
several groups have independently postulated fossil disk accretion models, 
\citep{cha00,alp01,mar01} only \citet{cha00}, CHN,  have provided a detailed,
time-dependent mass-transfer mechanism and, as a result, our analysis focusses
on their model.  In fact, we have shown that the CHN model contains inherent
inconsistencies and, ultimately, when measured against observations and
realistic phyiscal assumptions, fails.  In section~\ref{opacity}, for example, 
we have shown that a realistic estimate of the disk opacity forces us to either
assume unusually strong neutron star magnetic fields, or abandon the CHN model
altogether.  Using the alpha-disk model of \citet{sha73}, we have shown that 
a Kramers opacity must be assumed for 
any fossil disk with a mass-transfer rate given by equation~(\ref{mdot}).  
This is in contrast with the analysis of CHN, where a non-physical opacity
parameter, $\Gamma(\kappa) = 7/6$, was assumed.  We can interpret
this result as being due to the fact that the magnetospheric boundary increases
with increasing field strength.  If the magnetic field, therefore, exceeds 
some minimum value, the inner edge of the fossil accretion disk necessarily 
truncates far enough away from the neutron star such that temperatures 
are low enough to ignore electron scattering.  Furthermore, we have 
shown that a neutron star in a Kramers disk $(\Gamma_{\rm Kr} = 1.25)$, can 
not spin down to the AXP range fast enough for $B_{12} \approx 1-30$.  We have
estimated the minimum pulsar field needed to give results consistent with CHN
to be $\sim 3.7 \times 10^{13}$ G.  Although seemingly unusual, there have
been recent discoveries of radio pulsars with inferred field strengths in this
range.  \citet{cam00} have reported the discovery of J$1119-6127$, a 1670-year
old pulsar apparently centered in a previously uncatalogued 
supernova remnant, with an inferred field strength $B=4.1 \times 
10^{13}$ G.  Additionally, the same group has reported the discovery of PSR
J$1814-1744$ with $\tau_{\rm c} = 85$ kyr and $B=5.5 \times 10^{13}$ G.  Note 
that such field strengths are quite close to the so-called magnetar critical
field, $B_{\rm c} \equiv m_{\rm e}^2 c^3/e\hbar = 4.4 \times 10^{13}$ G.  
It has been suggested that, at $B > B_{\rm c}$, QED processes such as photon 
splitting may inhibit pair-production cascades near the magnetic poles and, 
neutron stars with field strengths in this range should be radio-quiet 
\citep{bar98}.  In fact, both J$1119-6127$ and J$1814-1744$ seem to be ordinary
radio pulsars and have no discernible X-ray emission.  Complicating matters
further, we note that 1814-1744 has spin properties quite similar to the AXP, 
1E $2259+586$.  Thus, it seems that the quantum demarcation line outlined 
above may be a bit fuzzy and it is not clear whether or not a neutron star 
in a fossil disk with $B_{12} \ga 3.7$ can be observed as a radio pulsar or 
AXP.  In any case, the thesis of CHN, namely ``ordinary'' neutron stars can 
produce AXPs, seems a bit strained at this point.

Another apparent inconsistency with the CHN model is their assumed propeller
torque, given by equation~(\ref{chn_prop}).  We have seen that, for rapid 
rotators, this model assumes that the neutron
star forces propelled material to be ejected from the magnetospheric boundary
with the same angular velocity as the star itself -- i.e. it must be spun up
to corotation.  After noting that such a propeller torque is somewhat 
inconsistent with historical precedent, we realize there is insufficient 
observational data on the propeller effect to immediately rule it out.  The 
exact nature of the torque is, however, crucial in determining the overall 
evolution of neutron stars in fossil disks, as was pointed out in 
section~\ref{prop}.  For example, if one assumes material is propelled from 
the magnetosphere with the angular velocity of a Kepler particle then an AXP 
will not be observed.  Instead, the neutron star will either be observed as a 
radio pulsar, or in the case of an extremely high magnetic field $(B \ga
10^{14}-10^{15}$ G$)$, a dim propeller.  As a test, we applied the CHN 
propeller torque to the well-studied HMXB, SMC X-1 and compared it to known 
models of angular momentum transfer, namely the energy and angular momentum 
propeller torques.  We found that, during the supergiant phase, the 
CHN model predicts a range of spin periods for SMC X-1 $\sim 2$ orders of 
magnitude different from the known period.  It seems that the propeller torque
employed by \citet{cha00} is simply too efficient to be consistent with 
observations.  A more robust comparison of propeller torque models would be
illuminating.  For example, one could examine the evolutionary history of all
HMXBs with known spin periods.  As there is a wide range of spin periods 
observed for these systems (especially the Be HMXBs), it is unknown whether or
not one propeller model can account for such variety.  In any case, our 
preliminary work seems to rule out the CHN propeller torque as physically 
unrealistic and, consequently, strikes a blow against fossil disk models in 
general.

Ultimately, only observations will definitively determine the true nature of 
anomalous X-ray pulsars.  In particular, optical and/or infrared studies of the
X-ray sources could help solve the mystery of AXPs once and for all.  A direct
detection of disk emission, for example, would strongly support the fossil disk
hypothesis.  Until now, the poor spatial resolution of X-ray telescopes have 
made precise positioning of AXPs difficult at best.  However, the recently 
deployed \emph{Chandra} observatory and \emph{Newton}XMM mission, 
with their much improved positioning abilities, should remedy this problem and 
help facilitate optical identifications.  In fact, the first possible optical 
counterpart to an AXP has recently been identified \citep{hul00}.  The 
proposed counterpart to 4U $0142 + 61$ has a measured flux ratio, $L_{\rm opt}/
L_{\rm x} \sim 10^{-3}$, an order of magnitude smaller than what has been 
predicted for fossil disk models \citep{per00}.  Although this measurement 
seems to, initially, lend more credence to the magnetar hypothesis, it should 
be noted that disk models can not be definitively ruled out either as the 
emission model of \citet{per00} sensitively depends on some poorly understood
parameters, such as disk size.


\clearpage

\begin{figure}
\figurenum{1}
\epsscale{0.8}
\plotone{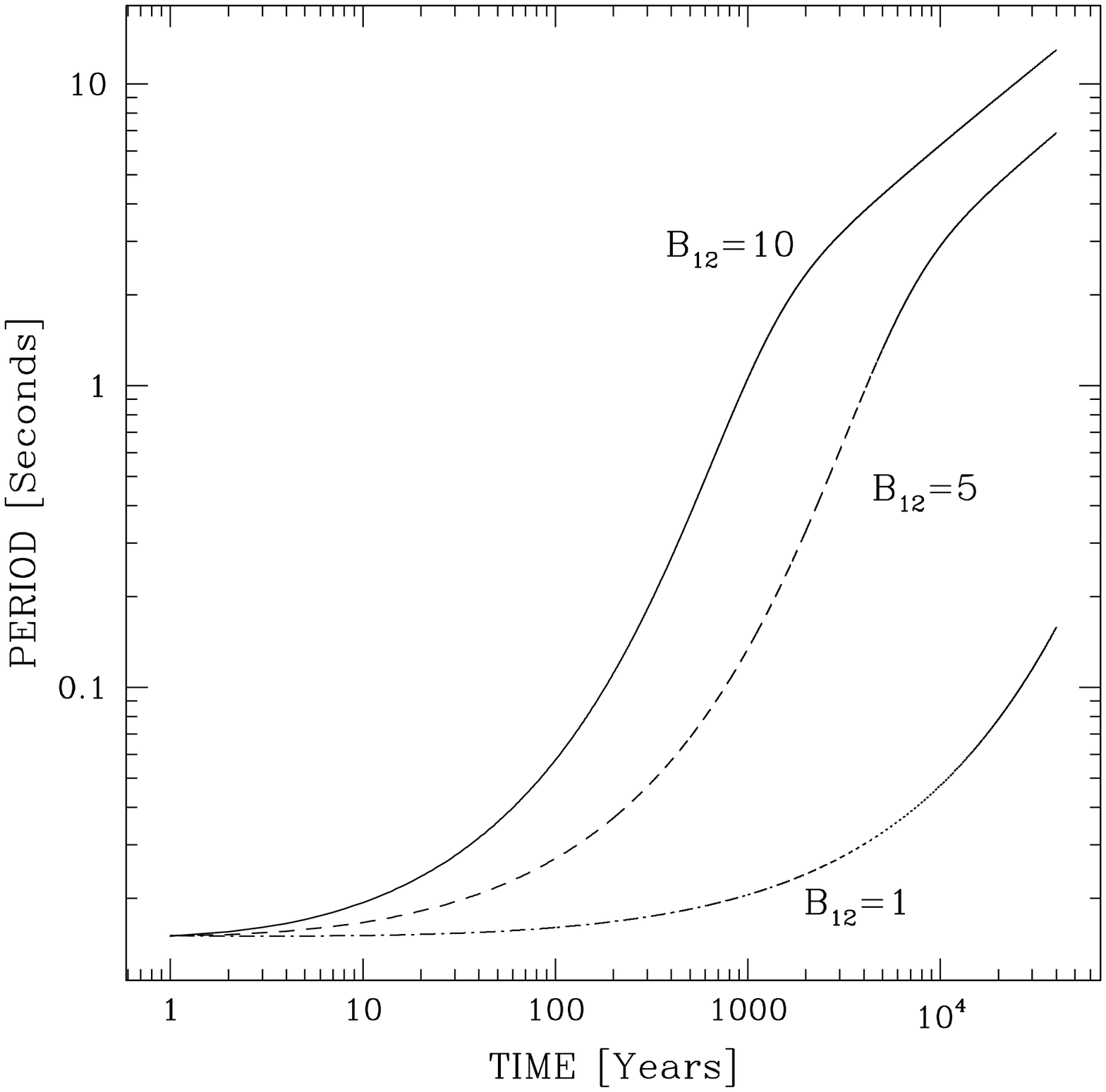}
\caption[]
{\footnotesize Time evolution of the pulsar spin period as a function of 
magnetic field strength for a neutron star accreting from a fossil disk. 
Assumed initial conditions are $P_0 = 15$ ms, $M_0 = 0.006 M_\sun$, and 
$\Gamma = 7/6$ (\emph{see text}).  The CHN ``efficient'' propeller torque, 
given by equation~(\ref{chn_prop}), is assumed.  We find that 
a magnetic field strength in the range $\sim 4-10 \times 10^{12}$ G
is needed for the neutron star to be observed as an AXP.  For weaker field
strengths, the neutron star can not spin down efficiently and is observed
as a radio pulsar.  When $B_{12} \ga 3.9$, a brief emitter phase is followed 
by an extensive propeller period.  The neutron star is observed as an AXP 
during the ``tracking'' phase of its evolution, i.e. when $\omega_{\rm s} \sim
1$.  In our model (as in CHN) the neutron star never enters an extended 
accretion period.  The AXP phase is assumed to exist, for the above initial 
conditions from the transtion time between propeller and tracking phases, 
$t_{\rm track} \sim 10^4$ yr, to the onset of ADAF, $2t_{\rm ADAF} \sim
38,000$ yr.  
}
\end{figure}

\clearpage

\begin{figure}
\figurenum{2}
\epsscale{.8}
\plotone{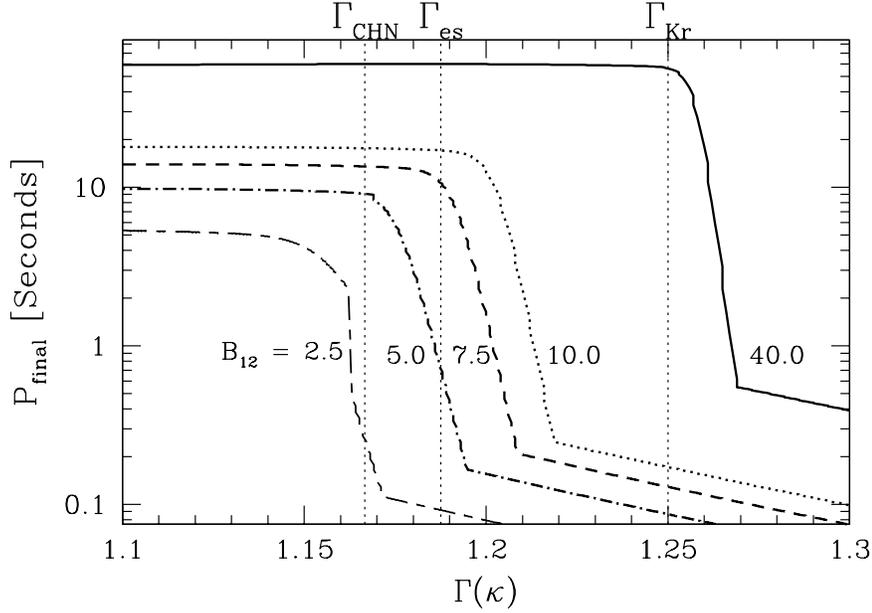}
\caption[]
{\footnotesize Final spin period of a neutron star embedded in a fossil disk 
as a function of opacity parameter, $\Gamma(\kappa)$ for various magnetic 
field strengths.  An efficient propeller torque is given by 
equation~(\ref{chn_prop}) and the ADAF transition time, $t_{\rm ADAF} = 
t_{\rm ADAF}(\Gamma)$ is given by equation~(\ref{ADAF}).  Five period-opacity
curves are shown for neutron star fields ranging from $2.5 \times 10^{12}$ G
to $4.0 \times 10^{13}$ G.  All curves show the same general trend and the 
existence of a cutoff opacity parameter, $\Gamma_{\rm c}$.  For a given field
strength, if $\Gamma \ga \Gamma_{\rm c}$, the neutron star can not sustain 
an extended propeller cycle and will, consequently, end its life as a radio
pulsar.  If electron scattering $(\Gamma_{\rm es}=1.1875)$ dominates in the 
disk, then the neutron star can be observed as an AXP provided $B_{12} \ga 
6.6$.  For a Kramers disk opacity $\Gamma_{\rm Kr}=1.25 \gg \Gamma_{\rm c}$
for all $B_{12} < 30$.  To observe an AXP in a Kramers disk, the minimum field
strength necessary is $\sim 3.7 \times 10^{13}$ Gauss.  In their analytic 
model, \citet{cha00} assumed $\Gamma = 7/6 \cong 1.167$.
}
\end{figure}

\clearpage

\begin{figure}
\figurenum{3}
\epsscale{.8}
\plotone{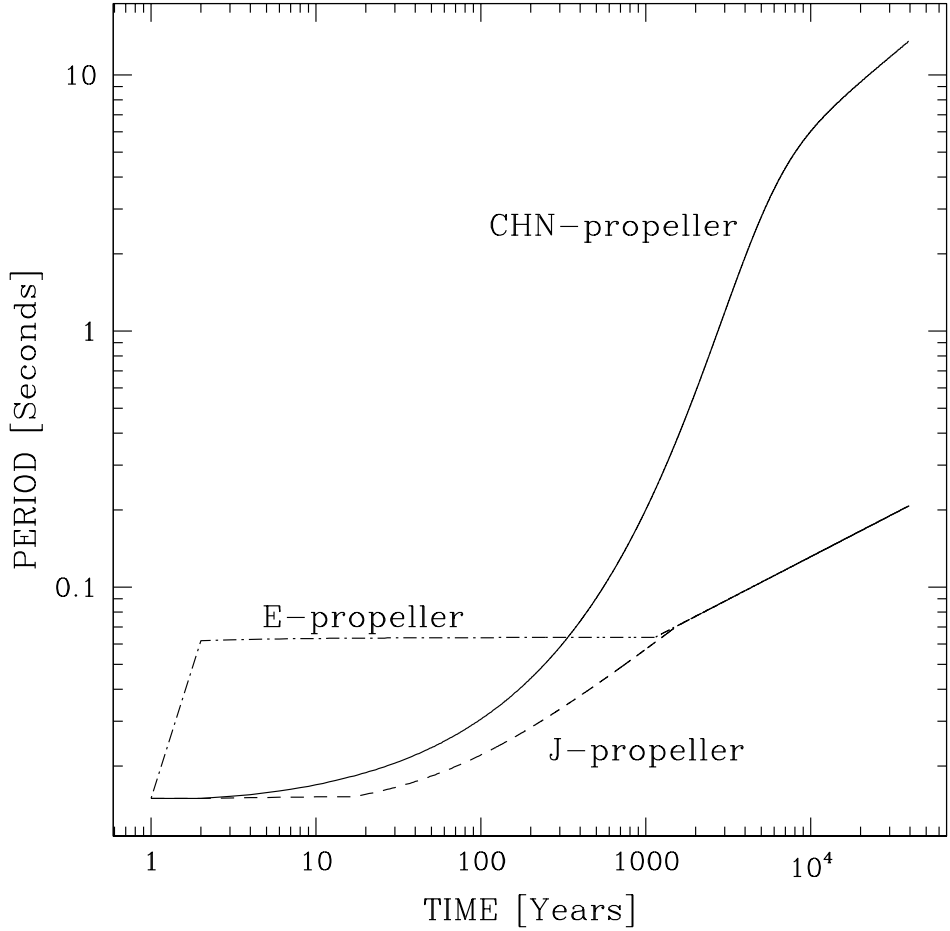}
\caption[]
{\footnotesize Spin evolution for a neutron star embedded in a fossil disk
for three different propeller torques.  Initial conditions for the various 
torques are otherwise fixed:$M_0 = 0.006 M_\sun$, $P_0 = 15$ ms, $B_0 = 7.5 
\times 10^{12}$ Gauss, and $\Gamma(\kappa) = 7/6$.  Employing a CHN propeller
torque yields, after a sharp propeller phase, an AXP in the tracking phase 
near $t \sim 1.5 \times 10^4$ yr.  The neutron star remains at 
quasi-equilibrium until $2t_{\rm ADAF} \sim 3.8 \times 10^4$ yr with a final
spin period, $13.5$ s.  For both the energy (E) and angular momentum (J) 
propellers, the neutron star never leaves the propeller phase of its evolution
and the final spin period is $\sim 0.2$ s.  In both cases, the neutron star is
observable as a radio pulsar.  The early evolutionary deviations for the energy
and angular momentum propellers are due to the lack of an emitter phase for the
former.  At later times, their spin evolution is consistent.
}
\end{figure}

\clearpage

\begin{figure}
\figurenum{4}
\epsscale{0.8}
\plotone{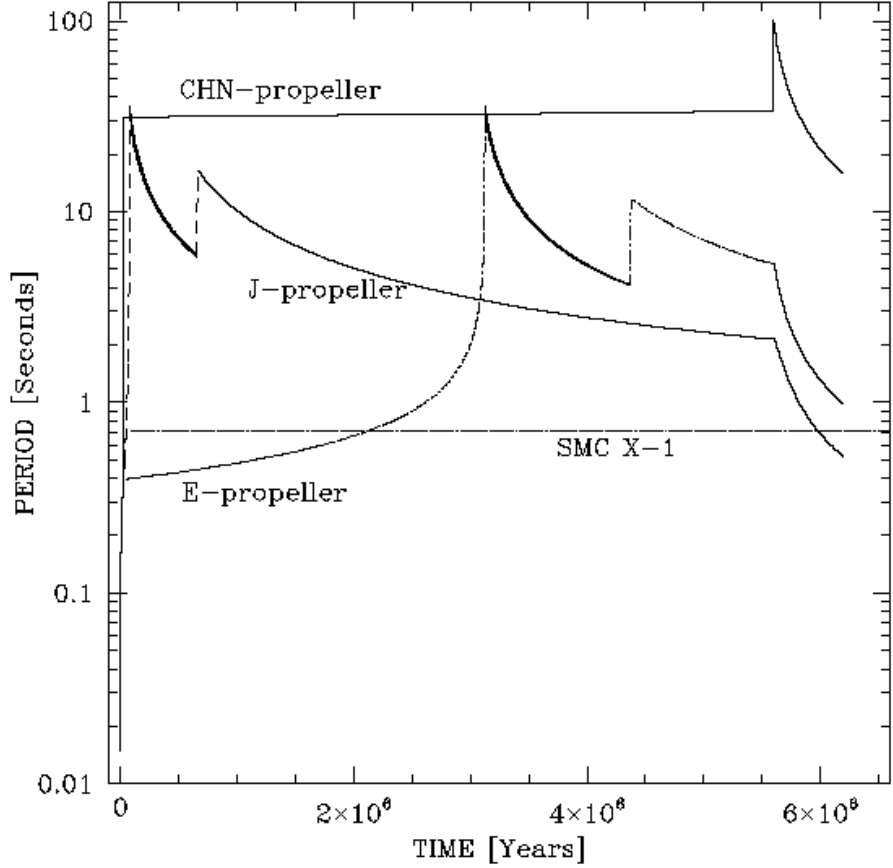}
\caption[]
{\footnotesize  Spin evolution models for SMC X-1 assuming various propeller
torque models.  The progenitor system, assumed to be an O9 V star and a newly
born neutron star, have the following initial properties: $M_1 = 20 M_\sun$, 
$R_1 = 10 R_\sun$, $B_0 = 10^{13}$ G, $P_0 = 15$ ms.  Mass is transfered by a 
radiation-driven wind during the (non-conservative) main sequence phase of the 
system's evolution.  At $t=5.6$ Myr, Hydrogen shell burning initiates in the
primary, instigating atmospheric Roche-lobe overflow.  If an energy propeller
is assumed, the calculated spin period during the supergiant phase ranges from 
$5.3$ s, at the onset of the heavy accretion cycle, to $0.99$ s at $t=6.2$ Myr.
For the angular momentum propeller, $P_{\rm f}$ ranges from $2.15$ s to $0.52$
s.  Assuming the efficient propeller torque of \citet{cha00} yields a neutron
star with a spin period ranging from $15.9$ s to $\sim 100$ s.  The observed
spin period of SMC X-1 $\cong 0.71$ s is illustrated for reference.
}
\end{figure}

\end{document}